\documentclass[a4paper,11pt]{article}
\usepackage{jcappub} 
\usepackage{lineno}
\usepackage{multirow}
\usepackage{array}
\usepackage{hyperref}
\usepackage[normalem]{ulem}
\usepackage[utf8]{inputenc}
\usepackage{amsfonts,amsmath,amssymb} 
\usepackage{graphicx,graphics,color}
\usepackage{tensor}
\usepackage[dvipsnames]{xcolor}
\usepackage{gensymb}
\usepackage{longtable}
\usepackage{bbding}
\usepackage{subfigure}
\usepackage{multirow}
\usepackage{float}
\usepackage{ulem}
\usepackage{soul}
\usepackage{verbatim}
\usepackage{mathtools}
\usepackage{comment}

\title{\boldmath Gravitational wave oscillations in Multi-Proca dark energy models}

\author[a]{Gabriel G\'omez,}
\author[b]{Jos\'e F. Rodr\'iguez}
 \affiliation[a]{Centro Multidisciplinario de F\'isica, Vicerrector\'ia de Investigaci\'on, Universidad Mayor,\\ Camino La Pir\'amide 5750,  Huechuraba, 8580745, Santiago, Chile}
 
 \affiliation[b]{Departamento de F\'isica, Universidad Antonio Nari\~no, Cra 3 Este \# 47A - 15, Bogot\'a D.C. 110231, Colombia}

\emailAdd{luis.gomezd@umayor.cl}
\emailAdd{jrodriguez154@uan.edu.co}

\abstract{Gravitational wave oscillations arise from the exchange of energy between the metric perturbations and additional tensor modes. This phenomenon can occur even when the extra degrees of freedom consist of a triplet of massive Abelian vector fields, as in Multi-Proca dark energy models. In this work, we study gravitational wave oscillations in this class of models minimally coupled to gravity with a general potential, allowing also for a kinetic coupling between the vector field and dark matter that can, in principle, enhance the modulation of gravitational wave amplitudes. After consistently solving the background dynamics, requiring the model parameters to reproduce a phase of late-time accelerated expansion, we assess the accuracy of commonly used analytical approximations and quantify the impact of gravitational wave amplitude modulation for current detectors (LIGO--Virgo) and future missions such as LISA. Although oscillations are present in these scenarios, we find that the effective mass scale (the mixing mass) governing the phenomenon is $m_g \sim \mu_A$, where $\mu_A$ is the (time-dependent) effective mass of the vector dark-energy field. 
Detectability of gravitational wave oscillations, however, requires $m_g \gg H_0$, which is in tension with the ultra-light masses typically needed to drive accelerated expansion  $\mu_A \sim H_0 \sim 10^{-33}\,\mathrm{eV}$. Therefore, if gravitational wave oscillations were to be detected at the corresponding frequencies, they could not be attributed to these classes of dark-energy models.}


\begin{document}
\maketitle
\flushbottom

\section{Introduction}
\label{sec:intro}

In 1918, Einstein predicted the existence of gravitational radiation, in close analogy with electromagnetic waves \cite{Einstein:1918btx}. Unlike their electromagnetic counterparts, gravitational waves (GWs) interact extremely weakly with matter, making their detection remarkably challenging. The discovery of the Hulse-Taylor binary pulsar \cite{Hulse:1974eb} provided indirect confirmation of gravitational radiation \cite{Weisberg:1981bh,Taylor:1982zz}, but it took nearly 100 years to achieve direct detection. The first detection by the LIGO-Virgo collaboration in 2015 \cite{LIGOScientific:2016aoc}, an event consistent with a binary black hole merger, marked the beginning of gravitational wave astronomy.

GWs offer a novel probe of fundamental physics and cosmology \cite{Lombriser:2015sxa,Koyama:2015vza,Ezquiaga:2017ekz,Baker:2017hug,Nishizawa:2017nef}. In particular, they provide an independent means to investigate the nature of dark energy, one of the central open problems in modern cosmology. Historically, the simplest explanation has been the cosmological constant $\Lambda$, interpreted as a vacuum energy density \cite{Peebles:1987ek,Weinberg:1988cp,Amendola:2015ksp}. 
However, recent joint analyses combining Baryon Acoustic Oscillation (BAO) measurements from the Dark Energy Spectroscopic Instrument (DESI) Data Release~2 \cite{DESI:2025fii}, cosmic microwave background (CMB) observations from \textit{Planck} \cite{Planck:2018vyg}, and Type~Ia supernova samples \cite{Scolnic:2021amr,Brout:2022vxf,DES:2024jxu} have provided increasing evidence that the dark energy component may evolve with time at low redshift \cite{DESI:2025zgx}. These indications add to existing tensions within the standard cosmological model, including the persistent $H_0$ tension \cite{Riess:2021jrx,Knox:2019rjx} and the long-standing $\sigma_8$ tension related to the amplitude of matter fluctuations \cite{Macaulay:2013swa,Battye:2014qga,Alam:2016hwk,Abbott:2017wau}.

Motivated by these anomalies, there has been renewed interest in dynamical dark energy scenarios. A well-studied class is provided by scalar-field models, commonly referred to as \textit{quintessence} \cite{Ratra:1987rm,Wetterich:1987fm}, in which cosmic acceleration is driven by a slowly evolving scalar field rolling along a sufficiently flat potential \cite{Caldwell:1997ii}. Beyond scalar-field descriptions, vector-like dark energy models have also emerged as viable and well-motivated alternatives
\cite{Koivisto:2007bp,Koivisto:2008xf,Mehrabi:2015lfa,DeFelice:2016yws,Rodriguez:2017wkg,Gomez:2019tbj,
Orjuela-Quintana:2020klr,deRham:2021efp,Gomez:2022okq,Cardona:2023gzq,
Pookkillath:2024ycd,Garcia-Serna:2025dhk}. Among these, a particularly distinctive class is provided by Multi-Proca theories
\cite{Armendariz-Picon:2004say,Koivisto:2012xm,Yamamoto:2012tq,Gomez:2020sfz,
Rodriguez-Benites:2023otm} (see also Ref.~\cite{BeltranJimenez:2016afo} for a generalized and systematic framework including derivative self-interactions). 
A key feature of vector dark energy models lies in the configuration of the background fields. A single vector field inherently introduces a preferred spatial direction, thereby breaking rotational invariance. In the absence of an internal global symmetry enforcing isotropy, this issue can be circumvented in two main ways:  either by considering purely time-like vector configurations, or by introducing a set of three identical, mutually orthogonal space-like vector fields—commonly referred to as a \emph{cosmic triad} \cite{Armendariz-Picon:2004say}. 

Since dark energy belongs to a hidden sector, its properties must be probed indirectly, primarily through gravitational interactions. In this context, the advent of GW astronomy opens a new observational window on the dark sector. Early insight into such possibilities dates back to the pioneering work of Gertsenshtein \cite{Gertsenshtein1960}, who showed that GWs can convert into electromagnetic waves while propagating through a constant magnetic field. In practice, however, the weakness of this interaction implies that observable conversion would require extremely large propagation distances and coherent magnetic fields, rendering detection unlikely. By contrast, if vector-like dark energy permeates the Universe, analogous mixing effects could be significantly enhanced. This naturally raises the question of whether GW propagation can be exploited as a probe of the fundamental nature of dark energy.

Remarkably, perturbations of Multi-Proca fields also support helicity--2 excitations. These additional tensor modes can couple directly to the metric perturbations, leading to distinctive modifications of GW propagation \cite{Caldwell:2016sut}. Such couplings may arise through friction, velocity, chirality, or mass-mixing terms in the linearized equations of motion \cite{BeltranJimenez:2019xxx}, altering the dispersion relation and enabling energy exchange between the two sectors.  These features are not unique to Multi-Proca fields; they also appear in bigravity models \cite{Max:2017flc, Brax:2017hxh}. In metric-affine theories, where gravity is modeled as a phenomenon arising from curvature, torsion, and nonmetricity, additional helicity-2 modes emerge and can source metric perturbations \cite{Aoki:2023sum}. Moreover, within gravitational effective field theory frameworks, massive spin-2 fields can consistently appear and interact with the metric perturbations in a model-independent manner; see, for example, \cite{Bordin:2018pca,Alberte:2019lnd, Kolb:2023dzp, Zhang:2025fck}. Interestingly, spectator sectors present in the early universe can efficiently source tensor perturbations already at linear order through mixing between the metric tensor and additional spin-2 degrees of freedom, as schematically described in \cite{Gorji:2023ziy} within an effective field theory framework. See also Refs.~\cite{Biagetti:2013kwa,Biagetti:2014asa,Fujita:2014oba} for earlier studies of spectator scalar fields, and Refs.~\cite{Maleknejad:2012fw,Gorji:2020vnh,Iarygina:2021bxq} for related works in the context of multi-vector field models.

A general framework for studying GW propagation in the presence of additional tensor degrees of freedom was developed in Refs.~\cite{BeltranJimenez:2019xxx,Ezquiaga:2021ler}. However, those analyses relied on a phenomenological parameterization of the mixing terms, leaving open the question of whether the assumed parameters are compatible with a cosmological background that genuinely produces late-time acceleration. A more self-consistent approach was adopted in Ref.~\cite{Caldwell:2018feo}, where GW amplitude modulation in the presence of a gauge-field dark energy component was studied alongside the background evolution. That analysis found that, for realistic models, the resulting amplitude oscillations remain far below observational sensitivity.

In this work, we focus on the Multi-Proca dark energy model introduced in Refs.~\cite{Armendariz-Picon:2004say,Gomez:2020sfz} and address the problem in a fully consistent manner. We first solve the cosmological background equations to reproduce the observed expansion history, including late-time acceleration. We then use the resulting background solution to study the coupled evolution of helicity--2 perturbations in both the metric and vector sectors, allowing us to quantify the exchange of energy between them.  Our goal is to evaluate the prospects for constraining such scenarios with current and future GW observations.

This paper is organized as follows. In section~\ref{sec_model}, we present the Multi-Proca model and discuss the background cosmological dynamics. In section~\ref{sec:Gws}, we derive the linearized perturbation equations and analyze the implications for GW propagation and observational constraints. Finally, we summarize our findings and outline future directions in section~\ref{sec:conclusions}.

\section{The model}\label{sec_model}

We begin by considering the action of a Multi-Proca theory minimally coupled to gravity, including a general self-interaction potential and a non-trivial coupling to the dark matter sector\footnote{Throughout this work we assume that radiation and baryons interact with the Proca fields only through gravity.}. The action reads \cite{Gomez:2020sfz}
\begin{equation}
S =\int \Bigr[\tfrac{M_{\rm Pl}^2}{2} R  - \tfrac{1}{4} F^a_{\mu\nu}F_a^{\mu\nu}- V(X)\\
+ f(X)\mathcal{L}_{\tilde{m}}(g_{\mu\nu},\psi) \Bigr]\sqrt{-g}d^4x, \label{eqn:action}
\end{equation}
where $M_{\rm Pl}^2=(8\pi G)^{-1/2}$ is the reduced Planck mass, $R$ is the Ricci scalar, $F^{a}_{\mu \nu}\equiv\partial_{\mu}A_{\nu}^{a}-\partial_{\nu}A_{\mu}^{a}$ denotes the field-strength tensor of the vector fields. The function $V$ is a general potential depending on $X\equiv-\tfrac{1}{2}\tensor{A}{_{a\mu}}\tensor{A}{^{a\mu}}$, while $\mathcal{L}_{\tilde{m}}(g_{\mu\nu},\psi)$ represents the Lagrangian of the dark matter field  $\psi$. Finally, $f(X)$ is the coupling function.

Varying the action with respect to the metric yields the field equations
\begin{equation}
    M_{\rm Pl}^2 G_{\mu\nu} = T_{\mu\nu} = K_{\mu\nu} + M_{\mu\nu},
    \label{eqn:Einstein}
\end{equation}
where $K_{\mu\nu}$ and $M_{\mu\nu}$ denote the contributions of the vector fields and the dark matter sector, respectively, to the total energy-momentum tensor of the Universe.  They are defined as
\begin{equation}
    K_{\mu\nu} = F_{a\nu \alpha } F^{a}{}_{\mu }{}^{\alpha } -  \tfrac{1}{4} F_{a\alpha \beta } F^{a\alpha \beta } g_{\mu \nu } -A_{a\nu } A^{a}{}_{\mu } V_{,X} - V g_{\mu \nu },    
    \label{eqn:TA}
\end{equation}
and
\begin{equation}
    M_{\mu\nu} = f\mathcal{M}_{\mu\nu} + f_{,X}\mathcal{L}_{\tilde{m}} A_{a\mu}\tensor{A}{^a_{\nu}} ,
    \label{eqn:Tm}
\end{equation}
with, 
\begin{equation}
\mathcal{M}_{\mu\nu} = -2 \frac{\partial \mathcal{L}_{\tilde{m}}}{\partial g^{\mu\nu}} + \mathcal{L}_{\tilde{m}}g_{\mu\nu},
    \label{eqn:Tmtilde}
\end{equation}
and $f_{,X}$ denotes differentiation with respect to $X$.
Similarly, variation with respect to the vector fields leads to the Proca equation
\begin{equation}
    \nabla_{\nu} \tensor{F}{_{a\mu}^{\nu}}  + \mu_A^2 A_{a\mu}  = 0,
    \label{eqn:AFE}
\end{equation}
where we have introduced the effective mass
\begin{equation}
    \mu^2_A \equiv   f_{,X}\mathcal{L}_{\tilde{m}} -  V_{,X}.
\end{equation}
The second term represents the contribution from the coupling to dark matter, which may be alternatively interpreted as generating an effective dark-matter pressure. Regardless of whether this contribution is viewed as modifying the vector sector or the matter sector, its physical effect is encoded in the combination above.

We consider a homogeneous and isotropic background described by the Friedmann-Lemaitre-Robertson-Walker (FLRW) metric, 
\begin{equation}
    ds^{2}=a^{2}(\eta)(-d\eta^{2}+\delta_{ij}dx^{i}dx^{j}),\label{eqn:metrin_flat}
\end{equation}
written in conformal time, with $d\eta=dt/a$. In theories involving vector fields that enjoy an internal symmetry in the action, as the one considered here, preserving the symmetries of the cosmological background requires either imposing an internal SO(3) symmetry or restricting to purely timelike vector configurations. In this work, we deliberately adopt the former option and focus on the spatial branch, taking $A_{0}^{a}=0$ as a physical assumption rather than as a gauge choice. Accordingly, we introduce a triad of mutually orthogonal, space-like vector fields with a common norm $\varphi$
\begin{equation}
    A_{\mu}^{a}\equiv \varphi(\eta) \delta_{\mu}^{a}.\label{eqn:triad}
\end{equation}
Here $a(\eta)$ denotes the scale factor. Varying the action on this background leads to the following field equations:
\begin{equation}
3M_{\rm Pl}^{2}\mathcal{H}^{2}=a^{2}f \rho_{\tilde{m}}+\frac{3 \varphi'^{2}}{2a^{2}}+a^{2} V = a^2 \rho_{T},
\label{eqn:Einstein00}
\end{equation}
\begin{equation}
M_{\rm Pl}^{2}(\mathcal{H}^{2}+2\mathcal{H}')=-\frac{\varphi'^{2}}{2a^{2}}+a^{2}V-\mu_{A}^{2}\varphi^{2} = -a^2 p_{T},
\label{eqn:Einstein11}
\end{equation}
\begin{equation}
\varphi''+ a^{2} \mu_{A}^{2}\varphi=0,\label{eqn:vector}
\end{equation}
where $\rho_T$ and $p_{T}$ are  the total energy density and isotropic pressure associated to both dark fluids, respectively.
Here, prime denotes derivatives with respect to the conformal time, $\mathcal{H}\equiv a'/a$ is the conformal Hubble parameter. 
Following Refs.~\cite{Brown:1992kc,Bertolami:2008ab}, we adopt the effective identification
$\mathcal{L}_{\tilde{m}} = -\rho_{\tilde{m}}$ for the cold dark matter sector at the level of the background equations. We emphasize that this choice is not intended to represent a complete variational formulation for a relativistic fluid, such as the Schutz--Sorkin action for perfect fluids \cite{Schutz:1977df,Brown:1992kc}. Rather, it should be understood as a practical macroscopic prescription commonly employed in theories where the matter Lagrangian appears explicitly in the field equations due to non-minimal couplings \cite{Harko:2010zi,Minazzoli:2012md,Avelino:2018rsb}.

The energy density and isotropic pressure associated with the vector fields, as measured by a comoving observer, are
\begin{equation}
    \rho_{A} = \frac{3}{2}\frac{\varphi'^2}{a^4} + V,\label{eqn:vector_density}
\end{equation}
\begin{equation}
    p_{A} = \frac{1}{2}\frac{\varphi'^2}{a^4} - V + \mu_A^2 \frac{\varphi^2}{a^2}.\label{eqn:vector_pressure}
\end{equation}
The corresponding continuity equations follow directly from the Bianchi identities and read
\begin{equation}
    \rho_{A}^{\prime}+3\mathcal{H}(\rho_{A}+P_{A})=\frac{3\varphi}{a^{2}}(\varphi'-\mathcal{H}\varphi)f_{,X}\rho_{\tilde{m}},
\end{equation}
\begin{equation}
\rho_{\tilde{m}}^{\prime}+3\mathcal{H}\rho_{\tilde{m}}=-\frac{3\varphi}{a^{2}}(\varphi'-\mathcal{H}\varphi)f_{,X}\rho_{\tilde{m}}.    
\end{equation}
These equations clearly show that the two components exchange energy through a dynamical interaction mediated by the coupling function. It is worth noticing that the general formulation presented here differs slightly from Ref.~\cite{Gomez:2020sfz}, which was written in cosmic time, but it is straightforward to verify that both descriptions are fully equivalent.


\subsection{Background evolution}

To solve for the background dynamics, all quantities are expressed in dimensionless form by normalizing to their present-day values. The set of background equations is written as a first-order system for the variables according to the state vector
$\{\psi, \chi, a, \mathcal{H}, \hat{\rho}_m\}$,
where the variables $\psi \equiv \varphi / M_{\mathrm{Pl}}$ and
$\chi \equiv \psi'$ represent, respectively, the rescaled vector field amplitude and its conformal-time derivative; and $\mathcal{\hat{H}}=\mathcal{H}/H_{0}$. Setting $a_{0}=1$, leads directly to $\hat{\mathcal{H}}_0 = 1$. All energy densities are normalized to the present critical density as
$\hat{\rho}_i = \rho_i / \rho_{\mathrm{crit},0}$, while the conformal time $\eta$ is expressed in units of $H_{0}^{-1}$.

We adopt two common forms for the self-interacting potential: an exponential potential and a power-law potential,
\begin{equation}
    \hat{V}_{1}(\hat{X})=\hat{V}_{01} e^{2\lambda \hat{X}},\quad \hat{V}_{2}(\hat{\tilde{X}})=\hat{V}_{02} (-2\hat{X})^{-\beta}.\label{eqn:potential}
\end{equation}
For the coupling function, we assume a power law coupling
\begin{equation}
     f(\hat{X})=(1+\alpha \hat{X})^{n},\label{eqn:coupling}
\end{equation}
where $\hat{X}=-\frac{3}{2}\left(\frac{\psi}{a}\right)^{2}$. Here, $\lambda$, $\beta$ and $\hat{V}_{0i}$ are free parameters of the potentials, while $\alpha$ and $n$ are the coupling parameters of the vector field. Since $f$ is a real-valued function, this implies the bound $\left(\frac{\psi}{a}\right)^{2}<\frac{2}{3\alpha}$ at all times.

The conditions at the present time are determined consistently from the Friedmann constraint,
\begin{equation}
 \mathcal{\hat{H}}^2 = a^{2} f \hat{\rho}_{m} + \frac{1}{2} \left(\frac{\chi}{a}\right)^{2} + a^{2}\hat{V}(\hat{X}).
\end{equation}
Considering the reference values\footnote{It means that the background solution will match the $\Lambda$CDM solution at the present time. Notice that this can generally be achieved with a non-vanishing, thought small, vector field, leading to a weak coupling to dark matter.} $\Omega_{m,0} = 0.32$ and $\Omega_{\mathrm{DE},0} = 0.68$, 
the scalar field amplitude today is fixed analytically. For the exponential potential, this is constrained to
\begin{equation}
\psi_0 = \sqrt{ -\frac{1}{3\lambda} 
\ln \!\left( \frac{\Omega_{\mathrm{DE},0} - \chi_0^{2}/2}{\hat{V}_{01}} \right) },\label{eq:exp_phi_0}
\end{equation}
which, in turn, establishes the bound $\hat{V}_{01}\gtrsim \Omega_{\rm DE,0}$ for $\lambda>0$, or the converse for $\lambda<0$, provided that the kinetic energy contribution today is very small, as expected.
For the power-law potential the constraint reads
\begin{equation}
\psi_0 = \sqrt{ \frac{1}{3\beta} 
\left( \frac{\Omega_{\mathrm{DE},0} - \chi_0^{2}/2}{\hat{V}_{02}} \right)^{-1/\beta} }.\label{eq:power_phi_0}
\end{equation}
As, $\chi_{0}\ll1$, we expect $\hat{V}_{02}\sim \Omega_{DE}$ and $\beta>0$.
All these choices ensure that $\psi_{0}$ is a real-valued quantity, condition that holds even for the uncoupled case. Notice that there exists an extra condition coming from the coupling sector that enforces the field amplitude to lie in the domain $\psi_{0}^{2}<\tfrac{2}{3\alpha}$. However, as $\psi_{0}$ is already fixed, this will set rather a plausible range for $\alpha$ values, ensuring thus consistency between free parameters.
Finally, the normalized matter density follows from
\begin{equation}
\hat{\rho}_{m,0} = \frac{\Omega_{m,0}}{f_0}, 
\qquad 
f_0 = (1 + \alpha \hat{X}_0)^n, \qquad \hat{X}_{0}= - \frac{3}{2} \psi_{0}^{2}.
\end{equation}
This means that for given values of the free coupling parameters $\alpha$ and $n$, the matter density today is fully determined. In absence of coupling, of course, it does not require information of the field amplitude $\psi_{0}$. This setup guarantees that the Friedmann constraint is exactly satisfied at the initial time by construction. To ensure the physical consistency of the numerical evolution,
the code continuously monitors the dimensionless Friedmann residual,
\begin{equation}
\Delta_F = 
\frac{\mathcal{\hat{H}}^2 - a^{2} (f\hat{\rho}_m + \tfrac{1}{2}\frac{\chi^{2}}{a^{4}} + \hat{V})}
{\mathcal{\hat{H}}^2},
\end{equation}
which remains at the level of machine precision throughout the integration.
The physical domain of parameters is restricted to
$\lambda > 1/3$, $\hat{V}_0 > \Omega_{\rm DE,0}$,
and coupling strengths  $|\alpha| \lesssim \mathcal{O}(1)$ with $n \in [-1,1]$, which are consistent with recent observational constraints \cite{Coelho:2025vmo}. We integrate backwards from today to the emission time, using Radau stiff integrator, ensuring regularity and numerical stability of the solutions.

From now on, we refer to \textit{model 1} as the case characterized by the exponential potential (see Eq.~\ref{eqn:potential}) together with the coupling function defined in Eq.~(\ref{eqn:coupling}), and to \textit{model 2} as the case with the power-law potential (see Eq.~\ref{eqn:potential}) featuring a weak coupling, as we will adopt small values for the coupling parameters. This classification encompasses the most representative scenarios within this class of vector-like dark energy models and allows us to extract general conclusions.

To characterize the background dynamics, it is useful to introduce the effective 
equation of state (EoS) $w_{\rm eff}$ and the vector dark energy EoS $w_{A}$, which 
provide a clear diagnostic of the dominant energy component and its evolution. They 
are defined as
\begin{equation}
    w_{\rm eff} = \frac{P_{T}}{\rho_{T}}, 
    \qquad 
    w_{A} = \frac{P_{A}}{\rho_{A}},
\end{equation}
and can be computed from Eqs.~(\ref{eqn:Einstein00})--(\ref{eqn:Einstein11}) for 
the total fluid, and from Eqs.~(\ref{eqn:vector_density})--(\ref{eqn:vector_pressure}) 
for the vector sector.

\begin{figure}[htbp]
\centering
\includegraphics[width=.45\textwidth]{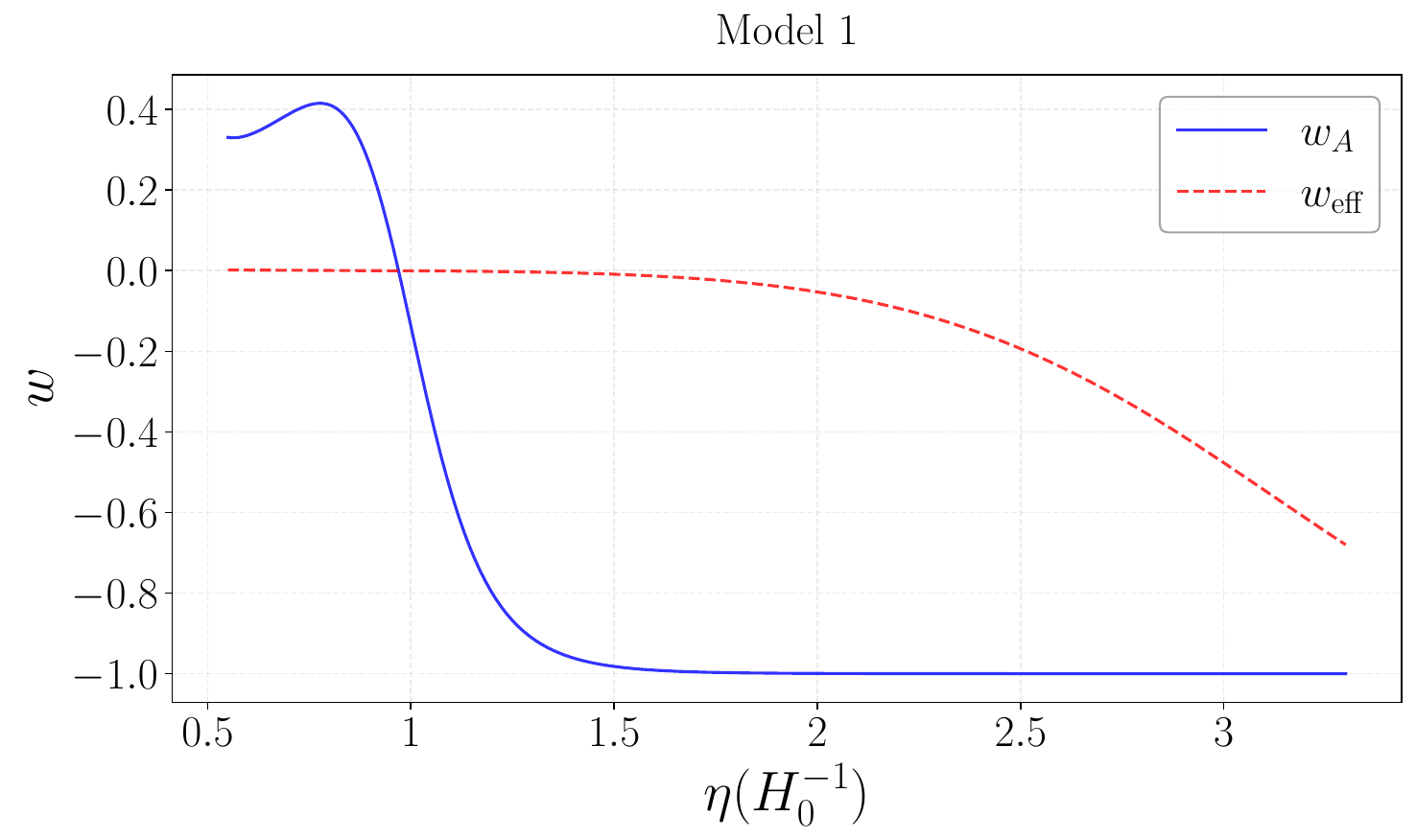}
\qquad
\includegraphics[width=.45\textwidth]{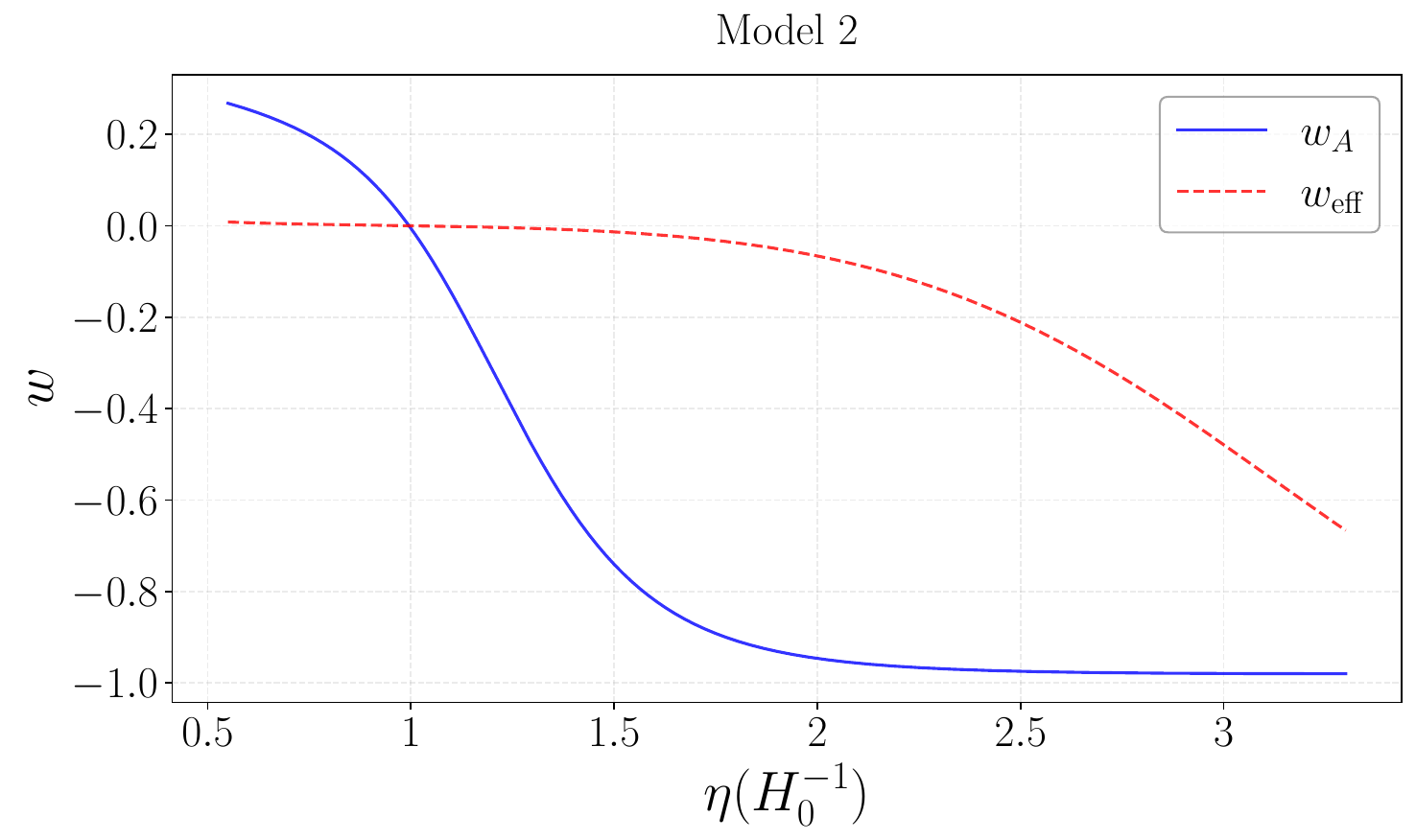}
\caption{Evolution of the effective EoS $w_{\rm eff}$ and the vector dark energy EoS $w_{A}$ as a function of conformal time $\eta$ for \textit{model 1} (strongly coupled exponential potential) in the \textit{left panel} and \textit{model 2} (weakly coupled power-law potential) in the \textit{right panel}. At late times, both models approach the de Sitter solution $w_{A} = -1$, while at intermediate stage, corresponding to the onset of the evolution, slight differences appear before reaching the radiation-like behavior $w_{A} = 1/3$. For model 1, this deviation is induced by the strong vector–matter coupling, which produces a sharp rise followed by a small bump. For model 1, we set the parameters $\alpha = 0.65$, $n = -0.3$, $\lambda = 0.34$, and $\hat{V}_{01} = 0.68001$. For model 2, we use $\alpha = -0.09$, $n = -0.009$, $\beta = -0.09$, and $\hat{V}_{02} = 0.690$. For both models we set $\chi_{0} = 5 \times 10^{-4}$, which uniquely determines the amplitude $\psi_{0}$ via Eqs.~(\ref{eq:exp_phi_0}) and (\ref{eq:power_phi_0}), respectively.}
\label{fig:EoS_evolution}
\end{figure}

Figure~\ref{fig:EoS_evolution} displays the evolution of $w_{\rm eff}$ and $w_{A}$ for 
the two models under consideration. At an intermediate stage\footnote{The emission time considered here corresponds to 
$z \approx 265$. Since matter–radiation equality occurs at $z \approx 3400$, the Universe 
is therefore well within the matter-dominated epoch at this stage, which is why we refer 
to it as an intermediate regime rather than early times, when radiation dominates the energy budget. The choice of such a high 
redshift is merely illustrative and serves to validate the consistency of the background evolution.}, both models exhibit small 
departures from the radiation-like value $w_{A}=1/3$, induced by the energy exchange 
term in the continuity equations. 
These deviations are more pronounced in \textit{model~1}, where the stronger 
coupling produces a sharper rise and a visible bump in $w_{A}$, whereas in 
\textit{model~2} the weaker coupling does not produce any observable signature.

As the universe expands and the interaction becomes subdominant, the dynamics of the vector field gradually approach the attractor behavior characteristic of the late-time de Sitter phase. Consequently, both models asymptotically converge to $w_{A}=-1$, signaling that the vector field \emph{successfully mimics a cosmological constant at late times}. $w_{\rm eff}$  follows the same trend, reflecting the eventual dominance of the dark energy sector over the total energy density. Although both models share the same late-time limit, their distinct early-time behavior may play an important role in determining the magnitude of the gravitational-wave amplitude modulation discussed in the following sections.

%


\section{Gravitational waves oscillations}\label{sec:Gws}
We investigate the propagation of GWs in the presence of a background vector field. To this end, we begin by perturbing the spacetime metric as
\begin{equation}
    ds^2 = g_{\mu\nu}dx^{\mu}dx^{\nu}=  (\bar{g}_{\mu\nu}+h_{\mu\nu})dx^{\mu}dx^{\nu} 
    \label{eqn:metric},
\end{equation}
where $\bar{g}_{\mu\nu}$ is the homogeneous and isotropic background metric introduced in Eq.~(\ref{eqn:metrin_flat}). In addition, the vector field is perturbed according to
\begin{equation}
    A_{a\mu} = \bar{A}_{a\mu} +  y_{a\mu}.
\end{equation}
The presence of both the internal index $a$ and the spacetime index $\mu$ allows the perturbations $y_{a\mu}$ to couple non-trivially to tensor modes of the metric. In other words, these perturbations support, and can modify, the propagation of helicity-2 or tensor degrees of freedom. From these definitions, the resulting linearized equations for the metric and vector-field perturbations can be found in Appendix \ref{app:TensorPert}. 

Since the background is homogeneous and isotropic, the perturbations of the field equations at linear order are decoupled according to their helicity \cite{Bardeen:1980kt}. The metric radiative degrees of freedom are traceless and transverse, thus we will restrict our study to helicity 2 degrees of freedom. The equation for the metric perturbations can be written in the following form
\begin{equation}
    \Box_s \delta^{ik}h_{kj}^{\rm TT} = -2M_{\rm Pl}^{-2} \bigl[\delta\tensor{K}{^{i}_j} + \delta\tensor{M}{^{i}_j}\bigr]^{\rm TT},
    \label{eqn:dGmunu1}
\end{equation}
where $\Box_s$ is the scalar wave operator (see e.g. \cite{1975ApJ...200..245T}),
\begin{equation}
    \Box_s \equiv (-\bar{g})^{-1/2}\partial_{\alpha}[(-\bar{g})^{-1/2}\bar{g}^{\alpha\beta}\partial_{\beta}],
    \label{eqn:Box_s}
\end{equation}
For simplicity, we consider perturbations propagating along the 
$z$-direction. Accordingly, the transverse–traceless metric perturbation can then be written as
\begin{equation}
    h_{ij}^{\rm TT} = a(\eta)^2
\begin{bmatrix}
    h_+ & h_{\times} & 0 \\
    h_{\times} & -h_+ & 0 \\
    0 &  0&  0
\end{bmatrix},
\label{eqn:ansatzh}
\end{equation}
and analogously the spatial components of the vector-field perturbation take the form
\begin{equation}
    y_{ai} = a(\eta)
\begin{bmatrix}
    y_+ & y_{\times} & 0 \\
    y_{\times} & -y_+ & 0 \\
    0 &  0&  0
\end{bmatrix}.
\label{eqn:ansatzy}
\end{equation}
Both $h_{ij}^{TT}$ and $y_{i}^{a}$ satisfy the transverse and traceless conditions: $\partial^{i}{h_{ij}}=h_{i}^{i}=0$ and $\delta_{a}^{i}\partial_{i}y_{j}^{a}=\delta_{a}^{i}y_{i}^{a}=0$. Thus, the tensor sector contains four dynamical degrees of freedom corresponding to the $+$ and $\times$ polarization states of the metric and vector fields. 

\medskip

To simplify the equations of motion, it is convenient to redefine the perturbation variables as
\begin{equation}
     h_P = \sqrt{2}\hat{h}_P/(aM_{\rm Pl}),\quad  y_P = \hat{y}_P/(\sqrt{2}a),
\end{equation}
with $P=\pm$. After transforming to Fourier space, it is convenient to group the tensor modes in the two-component column vector
$ \vec{\Phi} = \begin{bmatrix} \tilde{h}_k\\ \tilde{y}_k \end{bmatrix} $. Using the background equations and the definitions above,
the coupled system for tensor perturbations can be written in the compact matrix form: 
\begin{equation}
    \biggl(\mathbf{I}\frac{d^2}{d\eta^2} + \mathbf{I}k^2 + \mathbf{f}\frac{d}{d\eta}+\mathbf{M}\biggr)\vec{\Phi} = 0,\label{eqn:system_coupled}
\end{equation}
where $\mathbf{I}$ is the $2\times 2$ matrix identity. The friction matrix is given by
\begin{equation}
    \mathbf{f}= \begin{bmatrix}
    0 & f_{12} \\
    f_{21}& 0 
\end{bmatrix}=
\begin{bmatrix}
0 &   \frac{2 \varphi'}{a M_{\rm Pl}}  \\
     -\frac{2 \varphi'}{a M_{\rm Pl}} & 0 
\end{bmatrix},
\label{eqn:fmatrix}
\end{equation}
and the mass matrix reads
\begin{equation}
    \mathbf{M}= \begin{bmatrix}
    m_{11}^{2} & m_{12}^{2} \\
    m_{21}^{2}& m_{22}^{2} 
\end{bmatrix}=
\begin{bmatrix}
    \frac{2 \varphi^2 }{M_{\rm Pl}^2}\bar{\mu}_A^2   - \mu_H^2 & - \frac{2 \varphi a  }{M_{\rm Pl}}\bar{\mu}_A^2 \\
    \frac{2 \mathcal{H}  \varphi'}{a M_{\rm Pl}}& a^2\bar{\mu}_A^2 
\end{bmatrix}.
\label{eqn:Mmatrix}
\end{equation}
Here we have defined $\mu_H^2 \equiv \frac{2 \varphi'^2}{a^2 M_{\rm Pl}^2} + \mathcal{H}'+ \mathcal{H}^2$, which encodes information of the standard friction term and kinetic energy of the vector field. The off-diagonal entries of $\mathbf{f}$ and  $\mathbf{M}$ are responsible, respectively, for the damping and oscillatory modulation of amplitude between the metric $h$ and the auxiliary tensor mode $y$, reflecting their gravitational mixing. To solve the coupled system consistently, we must first solve the background 
dynamics, which induces an explicit time dependence in the coefficients of 
Eq.~(\ref{eqn:system_coupled}).

The system cannot be diagonalized in general because $\mathbf{f}$ and  $\mathbf{M}$ do not commute, and can not be solved analytically because of the time-dependence coefficients\footnote{Given the antisymmetric structure of the friction matrix, the propagation equation can be cast into a compact form via the field redefinition $\vec{\Phi}=\mathbf{S}\vec{\Psi}$, where $\mathbf{S}$ is invertible by construction. The system then reduces to
\[
\left(\mathbf{I}\frac{d^{2}}{d\eta^{2}}+\mathbf{I}k^{2}+\mathbf{M}_{\rm eff}\right)\vec{\Psi}=0,
\] with the effective mixing mass, given by $\mathbf{M_{\rm eff}}=\mathbf{S}^{-1}\mathbf{M}\mathbf{S}-\frac{1}{2}\mathbf{f}^{\prime}-\frac{1}{4}\mathbf{f}^{2}$, encapsulates not only the original mass terms but also the effects induced by the friction matrix.}. However, both eigenmodes propagate with the same group velocity, ensuring that the tensor modes remain coherent during propagation.

The matrices $\mathbf{f}$ and $\mathbf{M}$ evolve adiabatically, i.e., their characteristic time is much longer than the typical period of the perturbations. Depending on the wavenumber $k$, the system admits either a Wentzel-Kramers-Brillouin (WKB) approximation (when $k$ is comparable to the matrix elements, see Appendix \ref{app:WKB}) or eikonal approximation (when $k$ is much larger, see Appendix \ref{app:eikonal}). It is therefore convenient to define a critical wavelength that marks the transition between the WKB and eikonal regimes for the mass term. Since the background scales are determined by the Hubble rate and the field mass is ultra-light ($m \lesssim 10^{-33}$ eV), a critical transition frequency is defined at $f_{\rm crit} \sim 10^{-18}$ Hz. 
\subsection{Oscillations of the Waveform}

In this subsection, we assess the accuracy of the WKB approximation by comparing it with the full numerical solution of the coupled perturbation system Eq.~(\ref{eqn:system_coupled}). The analysis is performed using the background solutions obtained in the previous section for \textit{model 1} and \textit{model 2}, where the model parameters are fixed to ensure a self-consistent late-time cosmology. 

Before proceeding, we clarify the analytical strategy adopted in this section. The coupled tensor system contains a nontrivial friction matrix which, through an appropriate field redefinition, can be absorbed into an effective mass matrix $\mathbf{M}_{\rm eff}$, as advertised earlier. In principle, the WKB formulas of Ref.~\cite{BeltranJimenez:2019xxx} should therefore be applied to $\mathbf{M}_{\rm eff}$. However, in the high-$k$ regime relevant for the WKB approximation, the friction contributions are subleading compared to the mass. Moreover, since the friction terms scale with first derivatives of the background fields, they are strongly suppressed during low-redshift GW propagation. Since the goal of this section is to obtain analytic insight into the parametric dependence of the oscillation frequency, we work directly with the mass matrix $\mathbf{M}$ as an accurate approximation. The full system, including friction effects, is retained in the numerical analysis presented below.

We refer the reader to Ref.~\cite{BeltranJimenez:2019xxx} for a comprehensive discussion of the WKB treatment of tensor mixing; in particular, Sec.~2.2.1 presents the derivation of the mixing induced by the mass matrix. As in bigravity, the observable modulation of the GW signal is governed by the mixing angle, defined as
\begin{equation}
    \tan^{2}\theta_{g}
    =
    \frac{m_{12}^{2} m_{21}^{2}}{m_{11}^{2}+ M^{2}\Delta/2}\, ,\label{eqn:tan_thetag}
\end{equation}
where $M^{2}\equiv m_{11}^{2}+m_{22}^{2}$ and
\(
\Delta=\sqrt{1-4\,\mathrm{det}(\mathbf{M})/M^{4}}-1
\)
follows the notation of Ref.~\cite{BeltranJimenez:2019xxx}.  
We evaluate the mixing angle numerically both at emission and at the present epoch in order to quantify its evolution during propagation, finding that
\begin{equation}
\textit{model 1}: \qquad 0.0503068\pi < \theta_{g} < 0.496988\pi,
\end{equation}
\begin{equation}
\textit{model 2}: \qquad 0.486369\pi < \theta_{g} < 0.499204\pi.
\end{equation}
The interval for \textit{model 1} interestingly overlaps with the region in which oscillatory features in bigravity could, in principle, be detectable by LISA, namely  
$0.05\pi < \theta_{g} < 0.45\pi$~\cite{LISACosmologyWorkingGroup:2019mwx}. In contrast, for \textit{model 2} the mixing angle is essentially frozen throughout the evolution. 

The origin of this behavior can be traced back to the different background dynamics induced by the two potentials. For \emph{model~1} (exponential potential), the vector amplitude evolves more dynamically during GW propagation, with its time derivative remaining non-negligible over extended intervals. This directly enhances the off-diagonal components of the mass matrix, $m_{12}^{2}$ and $m_{21}^{2}$, which depend on both the field amplitude and its derivatives, particularly near the emission time. As a consequence, $\theta_g$ can span a broader range. By contrast, in \emph{model~2} (power-law case), the background evolution is essentially monotonic and rapidly approaches an attractor solution, leading to a strong suppression of the mixing terms
and a much narrower evolution of $\theta_{g}$. Finally, since both models are constructed to reproduce a $\Lambda$CDM background at the present epoch, their mixing angles necessarily converge to very similar values today; any residual differences at $z=0$ arise solely from the distinct initial conditions specified in Eqs.~(\ref{eq:exp_phi_0}) and (\ref{eq:power_phi_0}).

The squared eigenfrequencies of the coupled tensor system can be written as (see Appendix \ref{app:WKB}) 
\begin{equation}
\theta_{\pm}^{2}
=
k^{2}
+\frac{1}{2}M^{2}
\pm
\frac{1}{2}
\sqrt{
\left(m_{22}^{2}-m_{11}^{2}\right)^{2}
+4\,m_{12}^{2}m_{21}^{2}
}\,.
\end{equation}
A central aspect of the propagation analysis is the identification of the characteristic oscillation frequency between the tensor modes. This frequency is controlled by the difference of the eigenfrequencies of the system. In the high-$k$ regime, $m_{ij}^{2}/k^{2}\ll 1$, they read
\begin{equation}
\delta\theta
\equiv
\theta_{+}-\theta_{-}
\simeq
\frac{1}{2k}
\sqrt{
\left(m_{22}^{2}-m_{11}^{2}\right)^{2}
+4\,m_{12}^{2}m_{21}^{2}
}\,.
\label{eq:dtheta_general}
\end{equation}
In the present model (see Eq.~(\ref{eqn:Mmatrix}) for definitions of $m_{ij}^{2}$), the off-diagonal contribution is proportional to $m_{12}^{2}m_{21}^{2}\propto \mathcal{H}\varphi'$, and is therefore dynamically suppressed during late-time propagation of gravitational waves, where the background fields evolve slowly. As a result, the mixing term provides only a small correction to the oscillation frequency, and one can safely approximate
\begin{equation}
\delta\theta \simeq \frac{\delta m^{2}}{2k}\,,
\qquad
\delta m^{2}\equiv m_{22}^{2}-m_{11}^{2}\,.
\end{equation}
Substituting the explicit expressions of the diagonal entries of the mass matrix, the oscillation frequency can be written as
\begin{equation}
\delta\theta \simeq \frac{\bar{\mu}_A^{\,2}}{2k}
\left(
a^{2}
-2\frac{\varphi^{2}}{M_{\rm Pl}^{2}}
+\frac{\mu_{H}^{2}}{\bar{\mu}_{A}^{2}}
\right),
\end{equation}
which evidences that the contribution from the off-diagonal mixing term enters as a small correction, suppressed by $\mathcal{H}\varphi'/\bar{\mu}_{A}^{2}$.
Consequently, in the high-$k$ limit, and according to the late-time dynamics, the oscillation frequency is controlled by the effective mass scale. It is therefore convenient to introduce the \emph{mixing mass} $m_{g}$ via
\begin{equation}
    \delta\theta \equiv \frac{m_{g}^{2}}{2k}
    \qquad\Rightarrow\qquad 
    m_{g} \sim \mu_{A}.
\end{equation}
Thus, the modulation oscillates with a characteristic frequency set by the mixing mass, whose magnitude is primarily controlled by the effective mass term in the mass matrix\footnote{This follows from the fact that, in the matrix component $m_{11}^{2}$, the term $\mu_{A}^{2}$ dominates over $\mu_{H}^{2}$, while $\bigl|2\varphi^{2}/(a^{2}M_{p}^{2})\bigr|\ll 1$ throughout the entire evolution from emission to today, as verified numerically.}. We can use the fact that the background evolution drives $\mu_{A}$ toward a value of order the present Hubble scale, $\mu_{A}\sim\mathcal{H}_{0}$, with the scale factor normalized such that $a_0=1$ today. Consequently  
\begin{equation}
    m_{g} \sim \mu_{A} \sim \mathcal{H}_{0}\approx 10^{-33}\rm{eV},
    \label{mixing_mass}
\end{equation}
which clearly shows that the mixing scale at the order of the present Hubble rate $\mathcal{H}_{0}$—precisely the mass scale expected for a DE scale field. In contrast to earlier studies, this small value of $m_{g}$ is not a free parameter but is instead fixed by the background solution,
which strongly constrains the viable dynamics of the tensor–mixing sector.


Next phenomenological aspect we want to examine is the usual assumption of  \emph{time-independent coefficients} for a given wavenumber $k$ in this model.  The numerical results show that for $k=100$ there exists a critical redshift  $z \gtrsim 3$ beyond which this assumption no longer holds. Interestingly, this regime approximately overlaps with the expected redshift range of standard sirens from supermassive black hole mergers, $z \sim \mathcal{O}(1-10)$, which is significantly larger than the typical LIGO/Virgo sources ($z \ll 1$).

\begin{figure}[htbp]
\centering
\includegraphics[width=.45\textwidth]{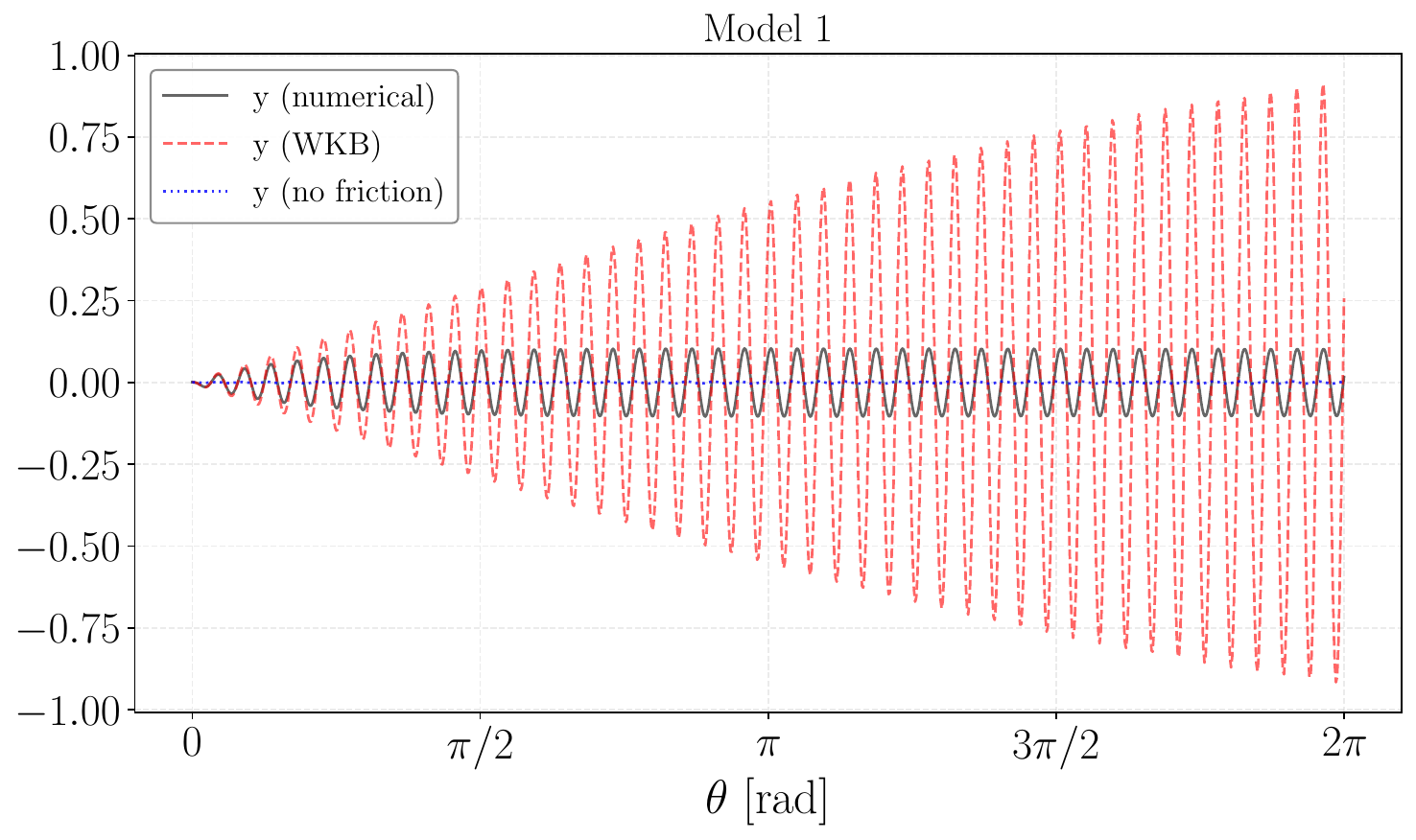}
\qquad
\includegraphics[width=.45\textwidth]{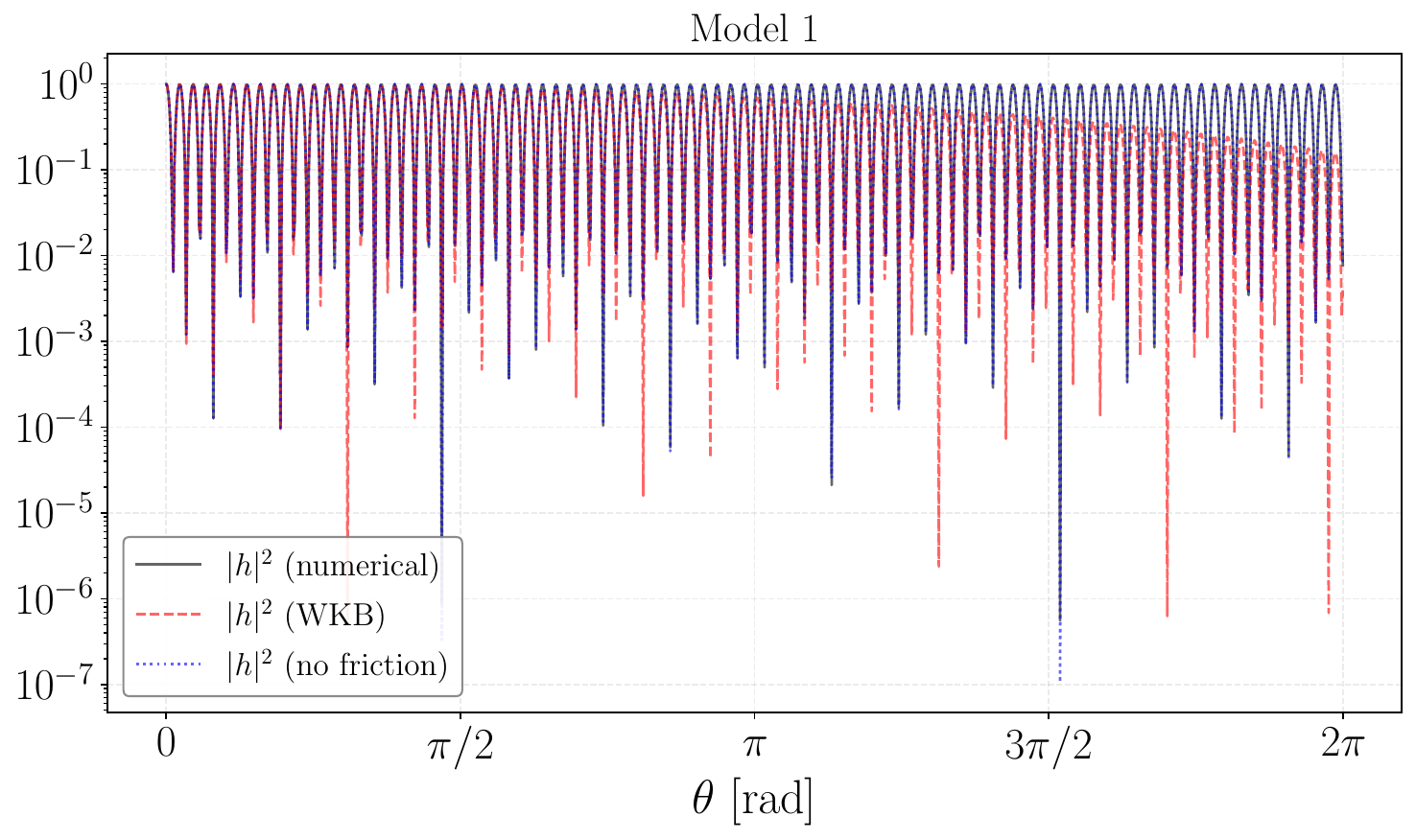}
\qquad
\includegraphics[width=.45\textwidth]{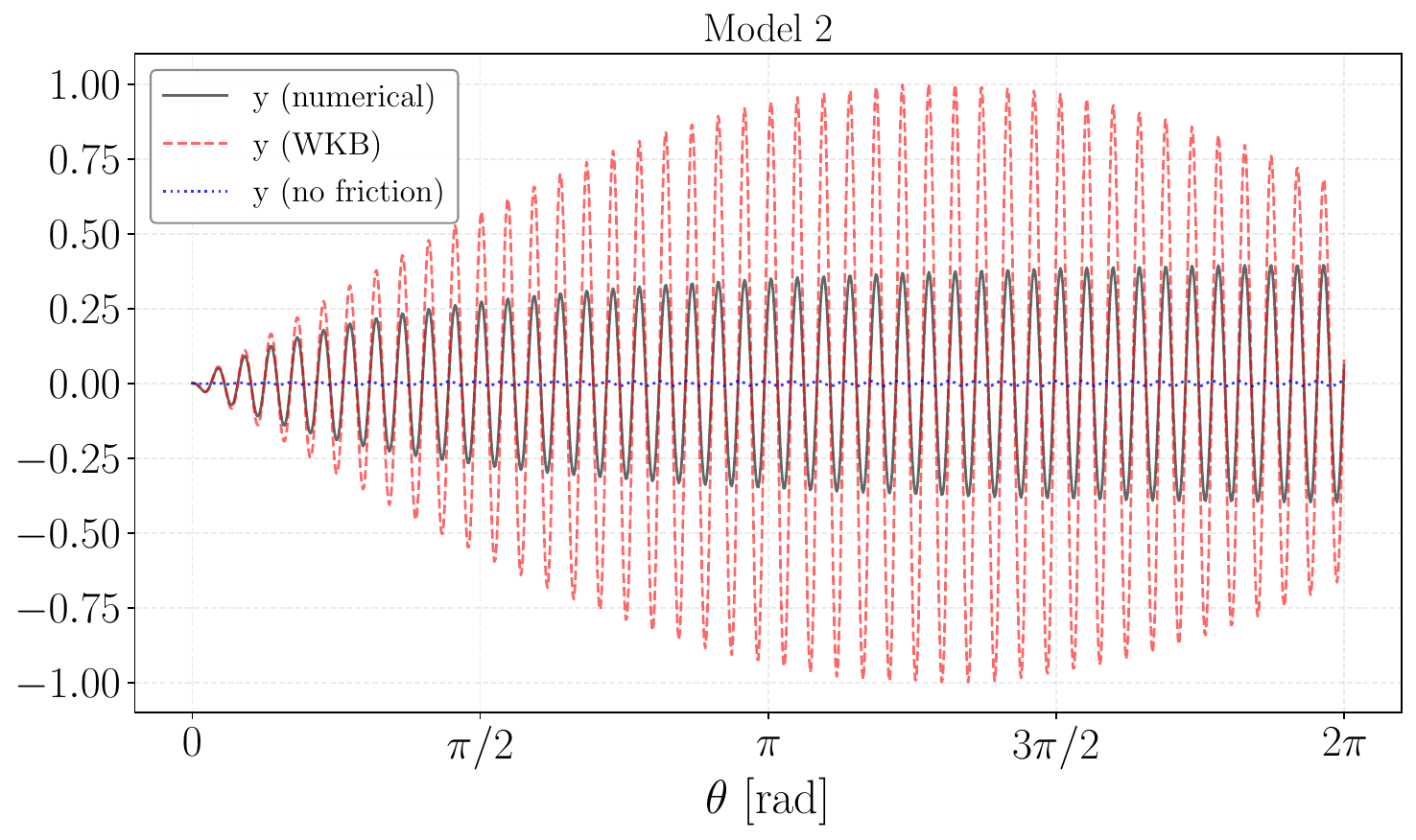}
\qquad
\includegraphics[width=.45\textwidth]{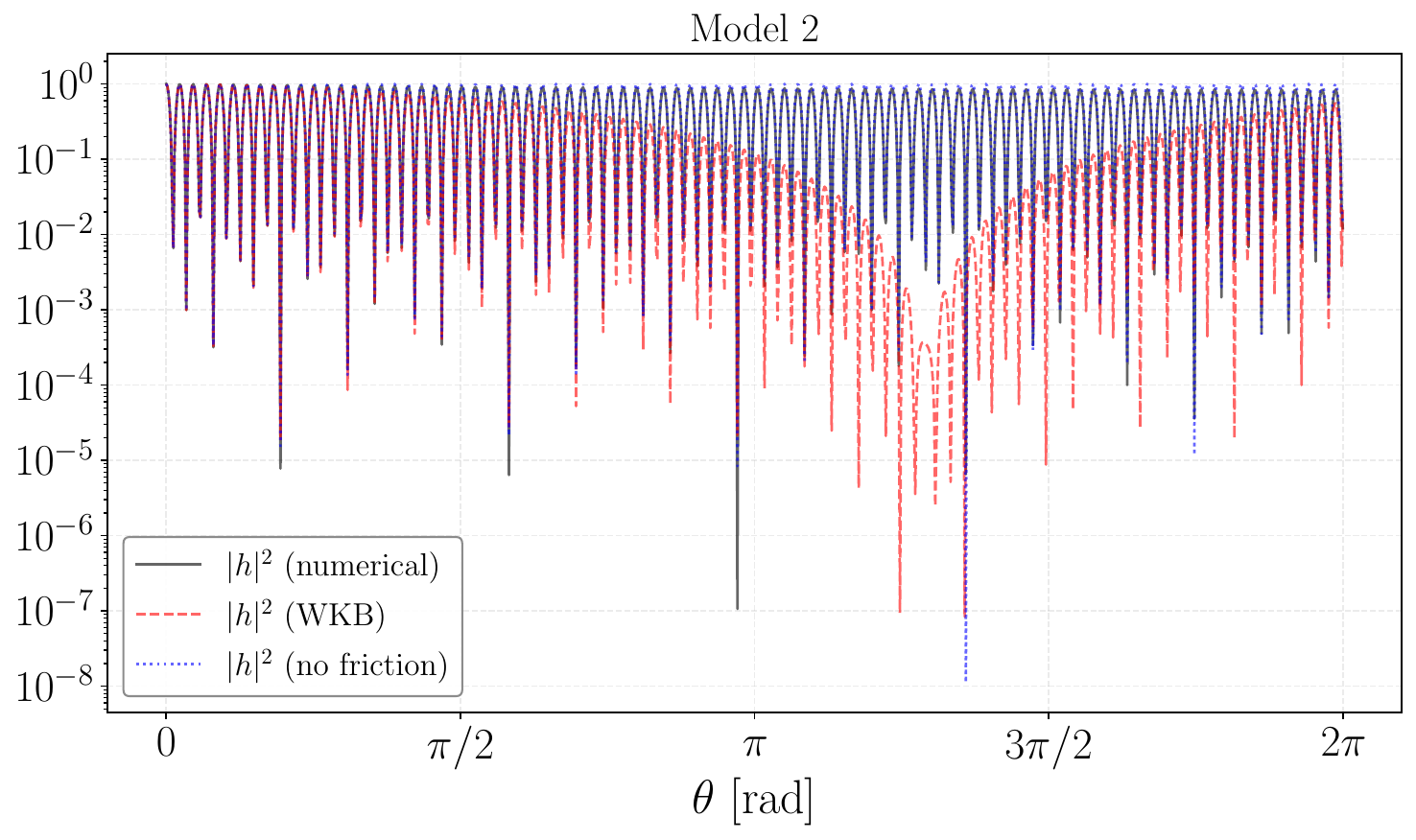}
\caption{Evolution of the auxiliary tensor mode $y$ (left) and the GW amplitude $|h|^{2}$ 
(right) for \textit{model~1} (top) and \textit{model~2} (bottom). Here, $\theta=2\pi(\eta-\eta_{e}/\Delta\eta)$, with $\Delta\eta=\eta_{0}-\eta_{e}$. 
We show the full numerical solution, the WKB approximation, and the result obtained 
by neglecting the friction term.  
The WKB method overestimates the growth of $y$, leading to an artificial suppression 
of the GW amplitude, while the full solution exhibits only mild damping.  The no inclusion of the friction term prevents significant growth of $y$, but keeping the GW amplitude essentially unchanged.  
In \textit{model~2}, the sign flip of $\mu_{H}^{2}$ produces an apparent reversal of 
mixing in the WKB approximation, but this effect does not occur in the numerical evolution, confirming that this feature is a methodological artefact. We have taken the following initial conditions at the emission time:
$h_{0}=1$, $y_{0}=0$, $h_{0}' = 10^{-3}$ and $y_{0}' = 10^{-3}$ and $k=100$.}
\label{fig:amplitudes_evolution}
\end{figure}

As a representative example for each model, Fig.~\ref{fig:amplitudes_evolution} shows 
the evolution of the auxiliary mode $y$ (left panels) and the GW amplitude $|h|^{2}$ 
(right panels) for \textit{model~1} (top) and \textit{model~2} (bottom).  
We display three cases of interest: the full numerical solution, the WKB approximation, 
and the solution obtained by neglecting the friction term.  
From these results it is clear that the WKB approximation \emph{overestimates} the 
growth of the dark auxiliary tensor mode (left panels), which in turn leads to an 
artificial suppression of the GW amplitude through the periodic conversion of $h$ 
into $y$ (right panels).

Increasing the wavenumber to $k=500$ produces a comparable net damping of the GW signal, albeit with faster oscillations. For this reason, it 
is not particularly informative to display even higher-$k$ results in this section. As expected for a secular effect, the discrepancy between the WKB and numerical solutions becomes increasingly significant for earlier emission times.

One may argue that the friction term $\mathbf{f}$ could be safely neglected, 
since its contribution is  subdominant at the present epoch. However, retaining this term is essential for higher-redshift emissions, where its magnitude grows substantially.  In general we find that $\mathbf{f} \sim \rho_{\rm Kin}^{1/2}/a(\eta)$, with $\rho_{\rm Kin} \sim \varphi'^2/a^4$.  Thus, at the emission time $\rho_{\rm Kin}$ may be several orders of magnitude smaller than the present critical density, $\rho_{{\rm crit},0}$, but still large enough to affect the early-time dynamics. Moreover, while the component $f_{12}$ increases sharply 
towards the past, $f_{21}$ exhibits the opposite trend (see Eq.~(\ref{eqn:fmatrix}) for definitions of $f_{ii}$) . This opposite evolution is responsible for the small amplitude attained by the vector perturbation $y$ (see left panel of Fig.~\ref{fig:amplitudes_evolution}). Despite this, the corresponding impact on the GW amplitude remains negligible (right panel), as the suppression generated by the friction term is efficiently counteracted by the background scaling. This behavior explains the very small amplitude associated with the tensor mode $y$, while the GW amplitude remains practically unaffected.

During the evolution, both effective masses $\mu_{A}^{2}$ and $\mu_{H}^{2}$ may undergo sign changes. 
This occurs most notably in \textit{model 2}, where $\mu_{H}^{2}$ switches sign and thus reverses the direction of the mixing-induced energy exchange between the tensor modes.  In the WKB treatment, this sign flip momentarily transforms the mixing from a damping regime, where mixing extracts energy from $h$, to an enhancing regime where the auxiliary mode injects energy back into the GW sector. However, this behavior is \emph{not} genuinely captured by the full numerical solution: the oscillation pattern remains monotonic, and no physical reversal of the amplitude transfer is observed.  This indicates that the apparent sign-flip effect is an artefact of the WKB approximation and does not correspond to an actual dynamical feature of the system.

The next question we wish to address is whether these effects are potentially observable within the frequency band of current and future GW detectors. This step is essential in order to establish a realistic physical scenario and to discard claims that may arise from regimes that are not experimentally accessible.

%

\subsection{A Realistic Scenario}

We inferred in the previous section that the modulation of the GW amplitude becomes more pronounced for high–redshift astrophysical sources. However, it remains unclear whether such effects can leave any appreciable imprint on the observable GW strain measured by ground–based or space–based interferometers. To address this question, we must translate our theoretical predictions into the characteristic frequency ranges of existing detectors: LIGO/Virgo operates at $f \sim 10~\mathrm{Hz}$, whereas LISA is sensitive to much lower frequencies, $f \sim 10^{-3}~\mathrm{Hz}$, each within their corresponding observational bands. The large LIGO/Virgo frequencies allow us to employ, in addition to the WKB treatment, the high–$k$ limit\footnote{In such a regime one has  
$k = \frac{2\pi f}{H_{0}} \sim 10^{20}$, which are far too large to be implemented in the numerical analysis of the previous section. This is precisely why several studies adopt a more moderate value, typically $k \sim 10^{3}$, when examining the dynamics of mixed helicity–2 modes (see e.g., \cite{BeltranJimenez:2019xxx,deCesare:2025ovv}).}.

The impact of a mass–type mixing on the observed GW amplitude can be encoded in a frequency and redshift–dependent \emph{transfer function}. Assuming that the effective mixing angle $\theta_{g}$ and mixing mass $m_{g}$ vary slowly along the line of sight, the ratio between the observed strain and the GR expectation is given by \cite{BeltranJimenez:2019xxx,LISACosmologyWorkingGroup:2019mwx}
\begin{equation}
\frac{|h(z,k)|}{|h_{\rm GR}|}
= \cos^{2}\theta_{g}
\Biggl[
1+\tan^{4}\theta_{g}
\\+2\tan^{2}\theta_{g}
\cos\biggl(
\int_{0}^{z} 
\frac{m_{g}^{2}}{2k}
\frac{dz'}{H(z')}
\biggr)
\Biggr]^{1/2},
\label{eq:transfer_massmix}
\end{equation}
which captures the leading oscillatory corrections to the waveform.  
The phase is proportional to $m_{g}^{2}/k$, and therefore highly suppressed at large frequencies. The transfer function depends on $\theta_{g}$, $m_{g}$, the source redshift, and the background expansion rate.  

In what follows, we evaluate the integral in Eq.~\eqref{eq:transfer_massmix} using the self–consistent Hubble rate $H(z)$ obtained from our Multi–Proca background solutions, rather than imposing a fiducial $\Lambda$CDM history. We expect, however, to obtain a very minor correction, that for this estimation is not crucial. More importantly, in order to have an observable effect, the ratio
$m_{g}^{2}/k$ must be of order $H_{0}$. Considering that $m_{g}\approx \mu_{A}(z)a(z)$, with $a(z)=\frac{1}{(1+z)}$, the oscillatory phase that controls the modulation
in the transfer function (see Eq.~\eqref{eq:transfer_massmix}) can be written as
\begin{equation}
\Phi(z,f)\equiv \frac{\mu_{A}^2}{2k}\,I(z),
\qquad
I(z)\equiv \int_0^z \frac{dz'}{(1+z')^2 H(z')}\,,
\label{eq:phase_def}
\end{equation}
where we have assumed that $\mu_{A}$ does not vary rapidly  so that it can be taken outside the integral. This is essentially true for low-$z$ sources. The condition for a sizable oscillatory effect is $\Phi\gtrsim 1$. Rearranging,
\begin{equation}
\Phi\sim 1 
\quad\Longrightarrow\quad
f \sim \frac{I(z)\,\mu_{A}^{2}}{4\pi}\,.
\label{eq:f_req}
\end{equation}
To obtain numerical estimates, we take $H_0\simeq 2.2\times10^{-18}~{\rm s}^{-1}$ and compute numerically $I(z)$. This is done for \textit{model 1} for the sake of example. The result shows that $I(z)$ saturates quickly with redshift: most of its contribution comes from the low–$z$ region ($z\lesssim1$) it being of order $1/H_0$, and for
$z\gtrsim 10$ one finds $I(z)\sim 10^{17}$~s (i.e., few $\times1/H_0$).
Inserting this into Eq.~\eqref{eq:f_req} gives
\begin{equation}
f_{\rm req} \sim 10^{-18}~{\rm Hz},
\end{equation}
an astronomically small value. This estimate shows that, for 
$m_g\sim 10^{-33}~{\rm eV}$, the phase $\Phi$ is completely negligible across the 
LIGO/LISA frequency bands and for any realistic source redshift: even at the lowest
PTA frequencies, $\Phi$ remains many orders of magnitude below unity.
Detectable oscillatory signatures of the type encoded in 
Eq.~\eqref{eq:transfer_massmix} would therefore require either much larger effective
mixing masses (well above the dark–energy scale) or access to unrealistically low GW
frequencies (well below the PTA range).

A crucial point in this analysis is that we have so far treated $\mu_{A}$ as a fixed parameter, even though it is intrinsically time–dependent. This approximation is adequate at low redshift, where the variation of $\mu_{A}(z)$ is mild. However, at higher redshift the effective mass may evolve significantly and can reach values much larger than its present-day value. Therefore, a consistent assessment of the accumulated oscillatory phase in Eq.~(\ref{eq:transfer_massmix}) must take into account the redshift evolution of $\mu_{A}$. Before turning to the numerical results, it is useful to examine a few limiting cases.

In the low-redshift limit ($z\ll 1$), the integrand in the cumulative phase becomes nearly constant to leading order, 
since $\mu_A\to H_0$ and $H(z)\to H_0$. Analytically this gives  $\Phi(z)\approx \frac{H_{0}z}{4\pi f}$, implying a linear but extremely suppressed growth at small $z$, 
consistent with the rapid numerical decay of the phase as $z\to0$.  
At high redshift, where $\mu_A^2(z)$ grows only mildly compared to the strong $(1+z)^2$ suppression,  the integral saturates and remains far below unity across a wide frequency range. Numerically, we find that even up to $z=200$ (unrealistic redshift sources), the cumulative phase remains 
$\Phi\ll1$ for representative frequencies spanning the PTA, LISA, and ground-based bands, with 
typical values $\Phi_{\max}\simeq 10^{-7}$ at $f=10^{-9}\,\mathrm{Hz}$, 
$\Phi_{\max}\simeq 10^{-13}$ at $f=10^{-3}\,\mathrm{Hz}$, and 
$\Phi_{\max}\simeq 10^{-18}$ at $f=10^{2}\,\mathrm{Hz}$.  

\begin{figure}[htbp]
\centering
\includegraphics[width=.5\textwidth]{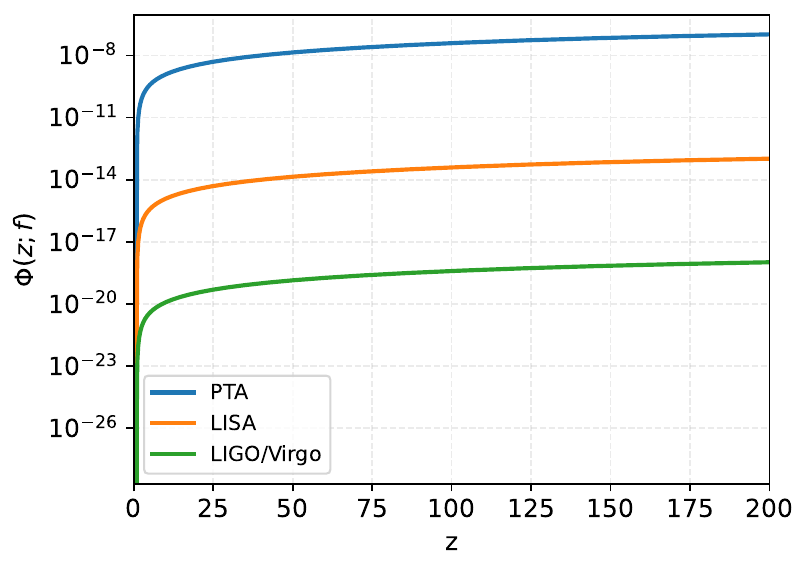}
\caption{Cumulative phase shift $\Phi(z;f)$ as a function of redshift for representative GW frequency bands (PTA: $f=10^{-9}\,\mathrm{Hz}$, LISA: $f=10^{-3}\,\mathrm{Hz}$, and LIGO/Virgo: $f=10^{2}\,\mathrm{Hz}$). The phase is computed
using the interpolated functions $\mu_{A}^{2}(z)$ and $H(z)$ for the \textit{model 1}. At low redshift the phase grows linearly but remains extremely small, while at high redshift the integral saturates, producing a plateau consistent with the weak redshift dependence of $\mu_{A}$. In all cases, the accumulated phase remains far below unity over the full range $z\le 200$.
}\label{fig:phase}
\end{figure}

These results confirm that, for late-universe propagation, mixing-induced phase accumulation is inefficient at both low and high redshift, making observable oscillatory effects highly suppressed. These findings are summarized in Fig.~\ref{fig:phase}, where the cumulative phase shift $\Phi(z;f)$ as a function of redshift for representative GW frequency bands (PTA, LISA, and LIGO/Virgo) is computed numerically, using the interpolated functions $\mu_{A}^{2}(z)$ and $H(z)$ from the background solution.

While oscillation–induced amplitude are intrinsically suppressed for astrophysical GW sources in these models, a particularly promising avenue for future work is the study of the stochastic GW background propagating through a dynamically evolving Multi–Proca sector in the early Universe, where the vector fields behave effectively as dark radiation. In the post-inflationary era, the vector amplitude can be many orders of magnitude larger than today, potentially allowing for significant amplification or suppression of the primordial tensor spectrum. Remarkably, even a relatively small energy density in a triplet of canonical $U(1)$ vector fields can leave observable imprints in the polarization B-mode pattern of the cosmic microwave background (CMB) \cite{Tishue:2021blv,Tishue:2022vwc}. This suggests that early-universe Multi–Proca dynamics may produce distinctive signatures accessible to next-generation probes of primordial gravitational waves, including the South Pole Telescope \cite{SPT-3G:2021eoc}, POLARBEAR \cite{Polarbear:2020lii}, and the Atacama Cosmology Telescope \cite{ACT:2025zrv}.


%
%

\section{Conclusions}\label{sec:conclusions}

In this work, we have investigated GW oscillations induced by the mixing between tensor metric perturbations and multiple vector fields, with particular emphasis on the class of Multi-Proca DE models. Although previous studies suggested that such \textit{dark spin-1 sectors} may imprint observable signatures on GW propagation, such as modulations of the waveform \cite{BeltranJimenez:2019xxx}, it remained unclear whether the proposed theoretical framework can simultaneously sustain a consistent cosmological background. Without this requirement, any predicted observational effect would lack physical reliability.

A key distinction of our analysis is that we neither impose a $\Lambda$CDM background nor rely on phenomenological parameterizations for the matrix coefficients. Instead, we solve self-consistently for the cosmological background, which in turn fixes the free parameters of the vector sector. Only  with this fully consistent setup can the predicted modulation of the GW amplitude be meaningfully and robustly quantified.

We have shown that GW amplitude modulation does occur in Multi-Proca–like dark energy models, exhibiting clear and well-defined oscillatory patterns in the waveform. These effects are more prominent for high-redshift sources, where neither the WKB approximation nor the frictionless limit fully captures the dynamics in the regime of low frequencies. Nonetheless, for low-redshift sources and high-frequencies signals, such as those relevant for LIGO/Virgo, these approximations remain fully adequate.

The central result of this study is that the modulation amplitude is severely constrained by the extremely small value of the mixing mass, $m_{g}\sim \mu_{A}\sim \mathcal{H}_{0}$, which is a direct consequence of enforcing a self-consistent background solution. This ultra-light mass regime coincides with that found in bigravity \cite{LISACosmologyWorkingGroup:2019mwx}, where the oscillatory factor $m_g^2/k$ cannot produce an $\mathcal{O}(1)$ modulation of the observed strain for any realistic detector frequency or source redshift. 

Our results are also consistent with those found for the \textit{gauge-essence} model, an $\mathrm{SU}(2)$ gauge-field dark energy scenario, where the suppression of the GW amplitude remains far below the sub-percent level even for sources located at very hight redshift \cite{Caldwell:2018feo}.

We therefore conclude that, if a modulation of the GW amplitude were ever detected by current ground-based interferometers or future space-based detectors, such an effect \emph{cannot} be attributed to multiple vector fields, either in Multi-Proca or $\mathrm{SU}(2)$ gauge-field realizations, within the class of vector-like dark energy models considered in this work. It remains an open question whether the inclusion of non-minimal couplings to gravity, capable of supporting self-accelerated solutions (see e.g., \cite{GallegoCadavid:2020dho,Garcia-Serna:2025dhk}), could modify this conclusion by allowing for a different mixing mass. Addressing this possibility lies beyond the scope of the present work and warrants a dedicated investigation.

Despite this pessimistic outlook for direct modulation effects, our analysis highlights an important opportunity: 
Spectator sectors present in the early universe can enhance the amplitude of the GW tensor spectrum already at linear order through mixing between the metric tensor and additional spin-2 degrees of freedom \cite{Gorji:2023ziy}. In the same spirit, the vacuum expectation value of a vector field can significantly affect the primordial stochastic GW signal even when the spectator sector contributes negligibly to the background energy density \cite{Iarygina:2021bxq}.
A related but distinct situation arises during radiation domination, where relic multi-Proca fields, while remaining subdominant, track the radiation background through their kinetic energy and can further modify the GW spectrum. This interaction may imprint characteristic signatures in the amplitude, tilt, and shape of the stochastic GW spectrum, consistent with the phenomenology discussed in \cite{Tishue:2021blv}. While our analysis does not explicitly employ an electric–magnetic decomposition, the presence of non-vanishing vector vacuum expectation values can, in more general settings, lead to parity-violating effects and the development of GW chirality, particularly in models with explicit dependence on magnetic-type components \cite{Tishue:2021blv}. In our framework, the tilt and oscillatory structure of the spectrum are directly linked to the mixing mass.
We emphasize that, although the general formalism for tensor mixing applies to both early and late-universe realizations and can, in principle, affect the GW spectrum at high frequencies, the allowed initial conditions and parameter space differ substantially between these cases. In particular, early-universe realizations must satisfy stringent constraints from Big Bang Nucleosynthesis and limits on extra relativistic degrees of freedom, commonly expressed in terms of $\Delta N_{\rm eff}$ \cite{Planck:2018vyg}. These constraints strongly restrict the energy density carried by additional fields and tensor modes at early times, whereas late-time dark-energy realizations are not subject to these bounds.



Finally, the benchmark parameter choices adopted in this work were intentionally selected to span a representative range of the parameter space and to test whether GW oscillations could become observable even in a strongly coupled vector--matter scenario, while remaining consistent with the background evolution. Since oscillatory signatures in the GW sector can not be detectable even in this strong-coupling regime, there is no phenomenological motivation to retain such large couplings (particularity for Model 1), as GW observations do not select or favor these values. One may therefore safely restrict attention to more natural regimes, e.g.\ $n\,\alpha \sim 10^{-5}$--$10^{-4}$, for which linear matter perturbations are expected to remain close to those of $\Lambda$CDM, in close analogy with coupled scalar-field models sharing the same coupling form \cite{Teixeira:2022sjr}.
In viable regions of parameter space with weak effective coupling, deviations in the growth of cosmic structures can be interpreted in terms of a mildly modified effective gravitational constant and additional source or friction terms that can suppress the growth of matter perturbations relative to the $\Lambda$CDM model \cite{Teixeira:2022sjr}. In addition, owing to the intrinsically spatial nature of the vacuum expectation value of the vector field, extra scalar degrees of freedom can be triggered at the level of linear perturbations. These effects may go beyond those typically encountered in scalar-field coupled dark energy models \cite{Koivisto:2005nr,Teixeira:2022sjr,daFonseca:2021imp} or in purely time-like vector-field coupled scenarios \cite{DeFelice:2020icf,Pookkillath:2024ycd,Cardona:2023gzq}. A dedicated scalar-perturbation analysis would be required to fully quantify these effects, and we therefore leave this investigation for future work.


\appendix
\section{Tensor perturbations}\label{app:TensorPert}
This Appendix details the derivation of the equations governing perturbation dynamics. We recall that barred quantities refer to those defined with respect to the background variables. The perturbed field equations take the following form:
\begin{equation}
    \delta \tensor{G}{^\mu_\nu} = M_{\rm Pl}^{-2}\delta\tensor{T}{^\mu_\nu}=M_{\rm Pl}^{-2}\bigl(\delta K^{\mu}{}_{\nu} + \delta M^{\mu}{}_{\nu}\bigr),
    \label{eqn:dGmunu0}
\end{equation}
and
\begin{multline}
    \bar{\nabla}_{\alpha }\delta F_{a}{}^{\alpha }{}_{\nu }- \tfrac{1}{2} \bar{F}_{a\nu }{}^{\alpha } \bar{\nabla}_{\alpha }h^{\beta }{}_{\beta }  + h^{\alpha \beta } \bar{\nabla}_{\beta }\bar{F}_{a\nu \alpha } + \bar{F}_{a\nu }{}^{\alpha } \bar{\nabla}_{\beta }h_{\alpha }{}^{\beta } + \bar{F}_{a}{}^{\alpha \beta } \bar{\nabla}_{\beta }h_{\nu \alpha } + y_{a\nu }  V_{,X} \\
+ \bar{A}_{a\nu } \bar{A}^{b\alpha } y_{b\alpha } V_{,XX} - \tfrac{1}{2} \bar{A}_{a\nu } \bar{A}_{b}{}^{\beta } \bar{A}^{b\alpha } h_{\alpha \beta } V_{,XX} + \bar{A}_{a\nu}\delta(f_{,X}\mathcal{L}_{\tilde{m}})+ y_{a\nu}\bar{f}_{,X}\bar{\mathcal{L}}_{\tilde{m}} = 0 \label{eqn:deltaF0},
\end{multline}
The perturbation of the mixed Einstein tensor is given by
\begin{multline}
    \delta G^{\mu}{}_{\nu} = - \bar{G}_{\nu }{}^{\alpha } h^{\mu }{}_{\alpha } + \tfrac{1}{2} \delta^{\mu }{}_{\nu } h^{\alpha \beta } \bar{R}_{\alpha \beta } -  \tfrac{1}{2} h^{\mu }{}_{\nu } \bar{R} -  \tfrac{1}{2} \bar{\nabla}_{\alpha }\bar{\nabla}^{\alpha }h^{\mu }{}_{\nu } + \frac{1}{2} \bar{\nabla}_{\alpha }\bar{\nabla}^{\mu }h_{\nu }{}^{\alpha } + \frac{1}{2} \bar{\nabla}_{\alpha }\bar{\nabla}_{\nu }h^{\mu \alpha } \\
    -  \frac{1}{2} \delta^{\mu }{}_{\nu } \bar{\nabla}_{\beta }\bar{\nabla}_{\alpha }h^{\alpha \beta }+ \frac{1}{2} \delta^{\mu }{}_{\nu } \bar{\nabla}_{\beta }\bar{\nabla}^{\beta }h^{\alpha }{}_{\alpha }-  \tfrac{1}{2} \bar{\nabla}^{\mu }\bar{\nabla}_{\nu }h^{\alpha }{}_{\alpha },
    \label{dGud}
\end{multline}
and the perturbation of the Faraday tensor is
\begin{equation}
    \delta F^{a}{}_{\mu\nu} = \bar{\nabla}_{\mu}y^{a}{}_{\nu} - \bar{\nabla}_{\nu}y^{a}{}_{\mu},
    \label{eqn:deltaF}
\end{equation}

The tensors constituting the perturbed energy-momentum tensor are:
\begin{multline}
    \delta \tensor{K}{^{\mu}_{\nu}}=  - \bar{F}_{a\nu }{}^{\beta } \bar{F}^{a\mu \alpha } h_{\alpha \beta } + \bar{F}_{a\alpha }{}^{\beta } \bar{F}^{a}{}_{\nu }{}^{\alpha } h^{\mu }{}_{\beta }  + \bar{F}^{a}{}_{\nu }{}^{\alpha } \delta F_{a}{}^{\mu }{}_{\alpha } + \bar{F}^{a\mu \alpha } \delta F_{a\nu \alpha } + \tfrac{1}{2} \delta^{\mu }{}_{\nu } \bar{F}_{a\alpha }{}^{\lambda } \bar{F}^{a\alpha \beta } h_{\beta \lambda }\\
    -  \frac{1}{2} \delta^{\mu }{}_{\nu } \bar{F}^{a\alpha \beta } \delta F_{a\alpha \beta } -  \bar{A}^{a}{}_{\nu } y_{a}{}^{\mu }  V_{,X}  -  \bar{A}^{a\mu } y_{a\nu } 
    V_{,X} +  \bar{A}^{a\alpha } y_{a\alpha } \delta^{\mu }{}_{\nu } V_{,X} -\tfrac{1}{2} \bar{A}_{a}{}^{\beta } \bar{A}^{a\alpha } \delta^{\mu }{}_{\nu } h_{\alpha \beta } V_{,X} \\
    + \bar{A}_{a}{}^{\alpha } \bar{A}^{a}{}_{\nu } h^{\mu }{}_{\alpha } V_{,X}  +  \bar{A}_{a\nu } \bar{A}^{a\mu } \bar{A}^{b\alpha } y_{b\alpha } V_{,XX}  - \tfrac{1}{2} \bar{A}_{a\nu } \bar{A}^{a\mu } \bar{A}_{b}{}^{\beta } \bar{A}^{b\alpha } h_{\alpha \beta } V_{,XX},
\end{multline}
\begin{multline}
    \delta \tensor{M}{^{\mu}_{\nu}} = \tensor{\bar{\mathcal{M}}}{^{\mu}_{\nu}} \delta f + \bar{f} \delta \tensor{\mathcal{M}}{^{\mu}_{\nu}}+\delta(f_{,X}\mathcal{L}_{\tilde{m}})\tensor{\bar{A}}{_a^i}\tensor{\bar{A}}{^a_j} \\
    + \bar{f}_{,X}\bar{\mathcal{L}}_{\tilde{m}}(\bar{A}^{a}{}_{\nu } y_{a}{}^{\mu } + \bar{A}^{a\mu } y_{a\nu } -  \bar{A}_{a}{}^{\alpha } \bar{A}^{a}{}_{\nu } h^{\mu }{}_{\alpha }),
\end{multline}
where $\delta(f_{,X}\mathcal{L}_{\tilde{m}})$ can be expressed in terms of the following variables:
\begin{equation}
    \delta f = f_{,X}\delta X, \quad \delta f_{,X} = f_{,XX}\delta X, 
\end{equation}
\begin{equation}
    \delta X = \tfrac{1}{2}\bar{A}^{a\mu}\bar{A}_{a}{}^{\nu}h_{\mu\nu} - A^{a\mu}y_{a\mu},
\end{equation}
\begin{equation}
    \delta \mathcal{L}_{\tilde{m}} = \tfrac{1}{2}(\bar{\mathcal{M}}^{\mu\nu}h_{\mu\nu}-\bar{\mathcal{L}}_{\tilde{m}}h^{\mu}{}_{\mu}).
\end{equation}

\section{WKB approximation}\label{app:WKB}
In this Appendix we briefly present the WKB approximation, which is discussed in detail in \cite{BeltranJimenez:2019xxx}. The WKB method provides an approximate solution for differential equations of the type shown in Eq. \eqref{eqn:system_coupled}. It is assumed that the phase varies rapidly, while the amplitude $\mathbf{\Phi}_n$, the friction matrix $\mathbf{f}$, and the mass matrix $\mathbf{M}$ vary slowly. The book-keeping parameter $\epsilon$ is introduced to keep track of the order of the approximation. Accordingly, Eq. \eqref{eqn:system_coupled} is rewritten as
\begin{equation}
    \biggl(\epsilon^2\,\mathbf{I}\frac{d^2}{d\eta^2} + \mathbf{C}k^2 + \epsilon\,\mathbf{f}\frac{d}{d\eta}+\mathbf{M}\biggr)\vec{\Phi} = 0,\label{eqn:system_coupled_epsilon}
\end{equation}
where $\mathbf{C}$ is the velocity matrix. For the case under study, this reduces to the identity matrix, implying that both perturbations propagate at the speed of light (see Appendix \ref{app:eikonal}).
The solution is obtained from the following ansatz 
\begin{equation}
\vec{\Phi} = \mathbf{E} \,\mathbf{G} \left( \vec{\Phi}_0 + \epsilon \vec{\Phi}_1 + \dots \right), \label{eqn:ansatz}
\end{equation}
where $\mathbf{G} \equiv e^{\frac{i}{\epsilon} \int \hat{\theta} d\eta}$, $\mathbf{E}$ is a matrix and  $\hat{\theta}$ is diagonal matrix. The solution is obtained by substituting the ansatz into Eq. \eqref{eqn:system_coupled_epsilon}, collecting terms of same order in $\epsilon$, and solving the resulting system of equations recursively. The solution is given by
\begin{equation}
\vec{\Phi}_{n} = \hat{\theta}^{-1/2} e^{-\int \mathbf{A}_{wkb} d\eta} \left( \vec{\mathbf{C}}_{n} + i \int e^{\int \mathbf{A}_{wkb} d\eta} \mathbf{B}_{wkb}^{-1} \vec{\mathbf{F}}_{n-1}^{wkb} d\eta \right), \label{eqn:WKB_gen}
\end{equation}
where the elements of the diagonal phase matrix are determined by the two roots $\theta_{1,2}$ of
\begin{equation}
\det \left[ \mathbf{C}\,k^2  + \mathbf{M} - \mathbf{I}\,\hat{\theta}^2 + i\mathbf{f}\,\hat{\theta} \right] = \det \left[\mathbf{W} - \mathbf{I}\,\hat{\theta}^2 + i\mathbf{f}\,\hat{\theta} \right]=0,
\end{equation}
with $\mathbf{W} = \mathbf{C}k^2 + \mathbf{M}$, $\mathbf{E}$ is the matrix of eigenvectors,
\begin{equation}
\mathbf{E} = \begin{pmatrix} 1 & -\frac{\mathbf{W}_{12} + i\mathbf{f}_{12}\theta_{2}}{\mathbf{W}_{11} - \theta_{j}^2 + i\mathbf{f}_{11}\theta_{j}} \\ -\frac{\mathbf{W}_{21} + i\mathbf{f}_{21}\theta_{1}}{\mathbf{W}_{22} - \theta_{2}^2 + i\mathbf{f}_{22}\theta_{1}} & 1 \end{pmatrix},
\end{equation}
the matrices $\mathbf{A}_{wkb}$, $\mathbf{B}_{wkb}$ are given by
\begin{equation}
\mathbf{A}_{wkb} = \mathbf{G}^{-1} \hat{\theta}^{1/2} (2\mathbf{E}\,\hat{\theta} - i\mathbf{f}\,\mathbf{E})^{-1} (2\mathbf{E}'\hat{\theta} - i\mathbf{f}\,\mathbf{E}' + \frac{i}{2}\mathbf{f}\,\mathbf{E}\,\hat{\theta}'\hat{\theta}^{-1}) \mathbf{G} \hat{\theta}^{-1/2},
\end{equation}
\begin{equation}
\mathbf{B}_{wkb} = (2\,\mathbf{E}\,\hat{\theta} - i\mathbf{f}\,\mathbf{E}) \mathbf{G} \,\hat{\theta}^{-1/2},
\end{equation}
the vector $\vec{\mathbf{F}}_{n-1}^{wkb}$ is
\begin{equation}
\vec{\mathbf{F}}_{n-1}^{wkb} = (\mathbf{E}\,\mathbf{G}\,\vec{\Phi}_{n-1}'' + 2\mathbf{E}'\mathbf{G}\,\vec{\Phi}_{n-1}' + \mathbf{E}''\mathbf{G}\,\vec{\Phi}_{n-1}),
\end{equation}
and $\vec{\mathbf{C}}_n$ are constant vectors. The leading order solution is,
\begin{equation}
\vec{\Phi}_{0} = \hat{\theta}^{-1/2} e^{-\int \mathbf{A}_{wkb} d\eta} \vec{\mathbf{C}}_{0}.
\end{equation}
\subsection{Mass Mixing}
In the case of negligible or vanishing friction, i.e., pure mass mixing, the two tensor modes ($h, t$) propagate and interact through a non-diagonal mass matrix $\mathbf{M}$:
\begin{equation}
\left[ \frac{d^2}{d\eta^2} + \begin{pmatrix} 1 & 0 \\ 0 & 1 \end{pmatrix} k^2 + \begin{pmatrix} m_{11}^2 & m_{12}^2 \\ m_{21}^2 & m_{22}^2 \end{pmatrix} \right] \begin{pmatrix} h \\ y \end{pmatrix} = 0.
\end{equation}
The associated eigenfrequencies $\theta_{1,2}$ for this system are given by
\begin{equation}
\theta_{\pm}^2 = k^2 + \frac{1}{2}M^2 \pm \frac{1}{2}\sqrt{M^4(1+\Delta)^2}.
\end{equation}
The definitions of $M$ and $\Delta$ can be found bellow Eq. \eqref{eqn:tan_thetag}. The matrix of eigenvectors $\mathbf{E}$ is
\begin{equation}
\mathbf{E} = \begin{pmatrix} 1 & -\frac{m_{12}^2}{k^2 + m_{11}^2 - \theta_2^2} \\ -\frac{m_{21}^2}{k^2 + m_{22}^2 - \theta_1^2} & 1 \end{pmatrix}.
\end{equation}
Therefore, the WKB solution \eqref{eqn:ansatz}, \eqref{eqn:WKB_gen} takes the following simpler form:
\begin{equation}
h(\eta)=\left[\frac{c_{1}}{\sqrt{\theta_{1}(\eta)}}+\frac{c_{2}}{\sqrt{\theta_{2}(\eta)}}\mathbf{E}_{12}(\eta)e^{i\int\delta\theta(\eta)d\eta}\right]e^{i\int\theta_{1}(\eta)d\eta},
\end{equation}
\begin{equation}
y(\eta)=\left[\frac{c_{2}}{\sqrt{\theta_{2}(\eta)}}+\frac{c_{1}}{\sqrt{\theta_{1}(\eta)}}\mathbf{E}_{21}(\eta)e^{-i\int\delta\theta(\eta)d\eta}\right]e^{i\int\theta_{2}(\eta)d\eta},
\end{equation}
where $\delta \theta \equiv \theta_+ - \theta_-$ and $c_1,c_2$ are constants given by the initial conditions. The form of the latter equations suggests the definition of the mixing angle given in \eqref{eqn:tan_thetag}.
\section{Large-$k$ approximation}\label{app:eikonal}
The large-$k$ approximation is a more restrictive case than the WKB method (see \cite{BeltranJimenez:2019xxx} for details): the wavenumber $k$ is large compared to the other parameters in the equations. Physically, this corresponds to the eikonal limit, representing the geometric optics limit for electromagnetic waves. This method is useful to obtain the characteristic velocities of the system. The eikonal approximation traces the characteristic surfaces of the wave-like system \eqref{eqn:system_coupled}, here rewritten as
\begin{equation}
\left[\mathbf{I}\frac{d^{2}}{d\eta^{2}}+\epsilon^{-2}\mathbf{C}\,k^{2}+\mathbf{f}\frac{d}{d\eta}+\mathbf{M}\right]\vec{\Phi}=0.
\end{equation}
Note that the friction and mass matrices are suppressed with respect to the velocity matrix. To solve the system in this approximation, the same ansatz \eqref{eqn:ansatz} and the same methodology of collecting terms according to the book-keeping parameter $\epsilon$ are used. The diagonal elements of $\hat{\theta}^2$ correspond to the eigenvalues of the velocity matrix $\mathbf{C}$ multiplied by $k^2$. Therefore, the characteristic speeds are $v^2_{1,2} = \hat{\theta}_{1,2}^2/k^2$. As previously noted, since $\mathbf{C}=\mathbf{I}$, the propagation is luminal. However, for the sake of generality, we retain the velocity matrix. The solution can be expressed as
\begin{equation}
\vec{\Phi}_{n}=\hat{\theta}^{-1/2}e^{-\int\mathbf{A}_{large-k}d\eta}\left(\vec{\mathbf{C}}_{n}+i\int e^{\int\mathbf{A}_{large-k}d\eta}\mathbf{B}_{large-k}^{-1}\vec{\mathbf{F}}_{n-1}^{large-k}d\eta\right),
\end{equation}
where
\begin{equation}
\mathbf{A}_{large-k} = \mathbf{G}^{-1}\hat{\theta}^{-1/2}\mathbf{E}^{-1} \left( 2\mathbf{E}'\hat{\theta} + \mathbf{E}\,\hat{\theta}' + \mathbf{f}\,\mathbf{E}\,\hat{\theta}  \right) \mathbf{G}\,\hat{\theta}^{1/2},
\end{equation}
\begin{equation}
\mathbf{B}_{large-k}=2\mathbf{E}\,\mathbf{G}\,\hat{\theta}^{1/2},
\end{equation}
\begin{equation}
\vec{\mathbf{F}}_{n-1}^{large-k}=\mathbf{E}\,\mathbf{G}\vec{\Phi}_{n-1}''+(2\mathbf{E}'+\mathbf{f}\,\mathbf{E})\mathbf{G}\vec{\Phi}_{n-1}'+(\mathbf{E}''+\mathbf{f}\,\mathbf{E}'+\mathbf{M}\,\mathbf{E})\mathbf{G}\vec{\Phi}_{n-1},
\end{equation}
and $\mathbf{E}$ is the matrix of eigenvectors
\begin{equation}
\mathbf{E}=\begin{pmatrix}1&-\frac{\mathbf{C}_{12}k^2}{\mathbf{C}_{11}k^2-\theta_{2}^{2}}\\ -\frac{\mathbf{C}_{21}k^2}{\mathbf{C}_{22}k^2-\theta_{1}^{2}}&1\end{pmatrix},
\end{equation}
The solutions are obtained recursively from
\begin{equation}
\vec{\Phi}_{0}=\hat{\theta}^{-1/2}e^{-\int\mathbf{A}_{large-k}d\eta}\vec{\mathbf{C}}_{0}.
\end{equation}

\acknowledgments

We thank Meng-Xiang for useful correspondence and for kindly providing Python notebooks for generating modified gravitational-wave waveforms. We are grateful to the anonymous reviewer for their insightful comments and constructive suggestions, which helped improve the manuscript.


\bibliographystyle{JHEP}

\bibliography{biblio}


\end{document}